\documentclass[12pt]{article}

\usepackage{amsmath}
\usepackage[dvips]{graphicx}
\usepackage{subfigure}

\newcommand\nn{\nonumber}
\newcommand\eq[1] {\begin{align} #1 \end{align}}
\newcommand\ga[1] {\begin{gather} #1 \end{gather}}

\newcommand{\br}[1]{\left( #1 \right)}
\newcommand{\brs}[1]{\left[ #1 \right]}
\newcommand{\brf}[1]{\left\{ #1 \right\}}
\newcommand{\brm}[1]{\left| #1 \right|}

\newcommand{\Sp}{\mbox{Sp}}
\newcommand{\Tr}{\mbox{Sp}}

\newcommand{\dd}[1]{{\hat #1}} % Dirac matrix conversion in russian style.

\newcommand{\GeV}{\mbox{GeV}}

\newcommand{\M} {{\cal M}} % Amplitudes.

\newcommand{\Moller} {{M{\o}ller }}

\begin{document}
% =========================================================================

\title{NNLO electroweak corrections for polarized \Moller scattering: one-loop insertions to boxes}

\author{
    A.~G. Aleksejevs\footnote{e-mail: aaleksejevs@grenfell.mun.ca}\\
    Memorial University of Newfoundland, Corner Brook, Canada\\
    \and
    S.~G. Barkanova\footnote{e-mail: svetlana.barkanova@acadiau.ca}\\
    Acadia University, Wolfville, Canada\\
    \and
    Yu. M.~Bystritskiy\footnote{e-mail: bystr@theor.jinr.ru}\\
    Joint Institute for Nuclear Research, Dubna, Russia\\
    \and
    \framebox{E. A.~Kuraev}\\
    Joint Institute for Nuclear Research, Dubna, Russia\\
    \and
    V. A.~Zykunov\footnote{e-mail: vladimir.zykunov@cern.ch}\\
    Belarusian State University of Transport, Gomel, Belarus
}

\date{\today}

\maketitle

\begin{abstract}
The paper discusses the two-loop (NNLO) electroweak radiative correc\-tions
to the parity-violating \Moller scattering asymmetry induced by
inser\-tions to boxes of electron and neutrino mass operators (fermion self-energies),
vertex functions and boson self-energies.
The results will be relevant to the ultra-precise $11~\GeV$ MOLLER experiment
planned at the Jefferson Laboratory, which will measure the weak charge of the electron and search for new physics.
The numerical estimations for the NNLO contribution to the cross section asymmetry are presented.
\end{abstract}


% =========================================================================
\section{Introduction}
\label{SectionIntroduction}
% =========================================================================

The M{\o}ller scattering \cite{Moeller:1932} with polarized electrons
has attracted active interest from both experimental and theoretical standpoints for several reasons.
It has allowed the high-precision determination of the electron-beam
polarization at SLC \cite{Swartz:1994yv}, SLAC \cite{Steiner:1998gf,Band:1997ee}, JLab \cite{Hauger:1999iv} and MIT-Bates \cite{Arrington:1992xu}
(and as a future prospect --- the ILC \cite{Alexander:2000bu}).
The polarized  M{\o}ller scattering can be an excellent tool in measuring parity-violating
weak interaction asymmetries \cite{Derman:1979zc}.
The first observation of Parity Violation (PV) in the M{\o}ller scattering was made
by the E-158 experiment at SLAC \cite{Kumar:1995ym,Kumar:2007zze,Anthony:2003ub},
which studied scattering of 45- to 48-GeV polarized electrons on the
unpolarized electrons of a hydrogen target.
It results at $Q^2 = -t = 0.026~\GeV^2$ for the observable parity-violating asymmetry
$A_{PV} = (1.31 \pm 0.14\ \mbox{(stat.)} \pm 0.10\ \mbox{(syst.)}) \times 10^{-7}$ \cite{Anthony:2005pm} which allowed one
of the most important parameters in the Standard Model (SM) -- the sine of the  Weinberg angle $\sin \theta_W $ -- to be determined
with accuracy of 0.5\,\%

The MOLLER (Measurement Of a Lepton Lepton Electroweak Reaction) experiment planned
at the Jefferson Lab aims to measure the parity-violating asymmetry in the scattering of
$11~\GeV$ longitudinally-polarized electrons from the atomic electrons in a liquid
hydrogen target with a combined stati\-sti\-cal and systematic
uncertainty of 2\,\%~\cite{vanOers:2010zz,Benesch:2011,Kumar:2009zzk,Benesch:2014bas}.
At such precision, any inconsistency with the Standard Model (SM) predictions will clearly
signal the new physics. However, a comprehensive analysis of radiative corrections
is needed before any conclusions can be made. Since MOLLER's stated precision
goal is significantly more ambitious than that of its predecessor E-158,
theoretical input for this measurement must include not only a full treatment of one-loop
(next-to-leading order, NLO) electroweak radiative corrections but also
two-loop corrections (next-to-next-leading order, NNLO).

The significant theoretical effort has been dedicated to one-loop radiative corrections already.
A short review of the references on that topic is done in \cite{Aleksejevs:2010ub,Aleksejevs:2012zz},
where  we  calculated a full set of the one-loop electroweak corrections (EWC)
both numerically with no simplifications using computer algebra packages
and by-hand in a compact form analytically free from nonphysical parameters,
and found the total relative correction to the observable asym\-metry to be close to $-70$\%.
It is possible that a large theoretical uncertainty in the prediction for the asymmetry may come
from two-loop corrections. One way to find some indication
of the size of higher-order contributions is to compare results that are expressed in
terms of quantities related to different renormalization schemes.
In \cite{Aleksejevs:2010nf}, we provided a tuned comparison between the result obtained with different
renormalization conditions, first within one scheme then between two schemes.
Our calculations in the on-shell and Constrained Differential Renormalization schemes show the difference of about 11\%,
which is comparable with the difference of 10\% between
$\rm \overline{MS}$ \cite{Czarnecki:1995fw} and the on-shell scheme  \cite{Petriello:2002wk}.
It is also worth noting that although two-loop corrections to the cross section may seem to be small,
it is much harder to estimate their scale and behavior for such a complicated observable as
the parity-violating asymmetry to be measured by the MOLLER experiment.

The two-loop EWC to the Born cross section ($\sim \M_0\M_0^+$) can be divided onto two classes:
$Q$-part induced by quadratic one-loop amplitudes $ \sim \M_1\M_1^+$,
and $T$-part -- the interference of Born and two-loop amplitudes $ \sim 2 \, \mbox{\rm Re} \br{\M_{0} \M_{2}^+}$
(here index $i$ in the amplitude $\M_i$ corresponds to the order of perturbation theory).
The $Q$-part was calculated exactly in \cite{Aleksejevs:2011de}
(using Feynman--t'Hooft gauge and the on-shell renormalization),
where we show that the $Q$-part is much higher than the planned experimental uncertainty of MOLLER,
i.e. the two-loop EWC are larger than was assumed in the past.
The large size of the $Q$-part demands detailed and consistent treatment of $T$-part, but this formidable task will require several stages. Our first step was to calculate the gauge-invariant  double boxes \cite{Aleksejevs:2012sq}.
In this paper we do the next step --
we consider the EWC arising from
the contribution of a wide class of the gauge-invariant Feynman
amplitudes of the box type with one-loop insertions:
fermion mass operators [or Fermion Self-Energies in Boxes (FSEB)],
vertex functions [or Vertices in Boxes (VB)],
and polarization of vacuum for bosons [or Boson Self-Energies in Boxes (BSEB)].

The paper is organized as follows.
We define the basic notations in Sect.~\ref{SectionCalculation} and present FSEB, VB, and BSEB in Sect.~\ref{sec.B}.
In Sect.~\ref{sec.NumbersAndConclusion}, we provide the numerical results for asymmetry for the kinematics conditions of the MOLLER
experiment and discuss work still to be done in the future.
In Appendix~\ref{appendix.FSEB}, the mass operators of electron and neutrino are presented.
In Appendix~\ref{appendix.VB}, we show the result for one-loop corrections to vertex functions for the case when only one fermion is on the mass shell.
In Appendix~\ref{appendix.BSEB}, we consider the polarization of vacuum for the virtual photon, $Z$- and $W$-boson.
The details of calculation of ultraviolet cut-off loop momenta integrals can be found in Appendix~\ref{appendix.Integrals}.

% vertex functions $V_{e e\gamma}$, $V_{e e Z}$, $V_{e \nu W}$ with one electron on mass shell;
% vacuum polarization operators $\Pi_{\gamma\gamma}$, $\Pi_{\gamma Z}$, $\Pi_{Z Z}$  and $\Pi_{W W}$.
% These quantities are inserted to the box type \Moller amplitude, producing the relevant chiral amplitudes.
% Corresponding contributions to the left-right asymmetry are calculated.
% The two loop vertex function of electron
% and two-loop polarization operator of the photon are considered in leading and next-to leading approximation.

% =========================================================================
\section{Basic notations}
\label{SectionCalculation}
% =========================================================================

We consider the process of electron-electron elastic scattering, i.e. \Moller process:
\eq{
e_-(p_1,\lambda_1)+e_-(p_2,\lambda_2) \to e_-(p_3,\lambda_3)+e_-(p_4,\lambda_4),
\label{1}
}
where $\lambda_i$ $(i=\overline{1,4})$ are the chiral states of initial and final electrons.
The kinematical invariants were defined in the standard way:
\eq{
s=(p_1+p_2)^2, \qquad t=(p_1-p_3)^2, \qquad u=(p_1-p_4)^2.
}
In the MOLLER experiment,  the expected beam energy is
$E_{\rm beam}=11~\GeV$, that is $s = 2m E_{\rm beam} \approx 0.01124~\GeV^2$,
where $m$ is the electron mass ($p_i^2=m^2$).
For the central region of MOLLER
(at $\theta \sim 90^\circ$ in center-of-mass system of initial electrons),  $-t\approx -u \approx s/2$
thus we can use an approximation that $s,\ |t|,\ |u| \gg m^2$.
Also, as for  MOLLER kinematics in central region $s,\ |t|,\ |u| \ll m_{Z,W}^2$
we neglect in following the terms of order ${\cal O}(s/m_{Z,W})$.

\begin{figure}
    \centering
    \mbox{
        \subfigure[]{\includegraphics[width=0.26\textwidth]{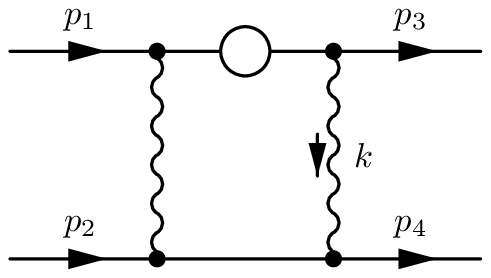}\label{fig.Fig3b}}
        \qquad
        \subfigure[]{\includegraphics[width=0.22\textwidth]{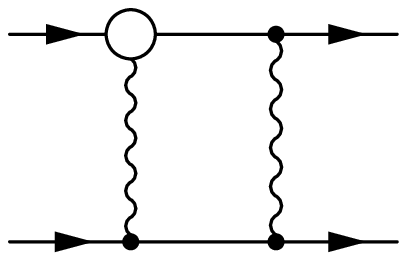}\label{fig.Fig3a}}
        \qquad
        \subfigure[]{\includegraphics[width=0.22\textwidth]{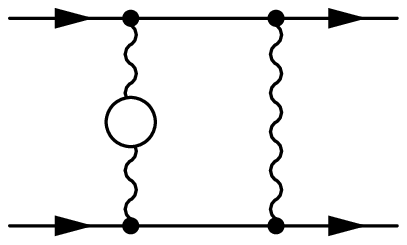}\label{fig.Fig3c}}
    }
    \caption{(a) Fermion self energies in boxes (FSEB), (b) boxes with vertices (VB), (c) boson self energies in boxes (BSEB).
            In this diagrams all wavy lines are assumed to be photons or $Z$-bosons. We also considered crossed box diagrams which can also contain $W$-bosons legs.}
    \label{fig.Boxeswithvertices}
\end{figure}

We consider the process (\ref{1}) in terms of chiral amplitudes $\M^{\lambda}$,
where $\lambda=\brf{\lambda_1 \lambda_2 \lambda_3 \lambda_4}$ is the chiral state of initial and final
electrons. The PV asymmetry to be measured by MOLLER is then defined as
\eq{
A=\frac{\brm{\M^{----}}^2-\brm{\M^{++++}}^2}{\sum_\lambda \brm{\M^{\lambda}}^2},
\qquad
\sum_\lambda \brm{\M^{\lambda}}^2=2(8\pi\alpha)^2\frac{s^4+t^4+u^4}{t^2u^2}.
}
In the Born approximation, this asymmetry has a form
\eq{
A^{(0)}=\frac{s}{2m_W^2}\frac{s^2tu}{s^4+t^4+u^4}\frac{a}{s_W^2}
}
proportional to
\eq{
a=1-4s_W^2.
}
Let us now recall that
$s_{W}\  (c_{W})$
is the sine (cosine) of the Weinberg angle expressed in terms of the $Z$- and $W$-boson
masses according to the Standard Model rules:
\eq{
s_{W}=\sqrt{1-c_{W}^2},\qquad
c_{W}=m_{W}/m_{Z}.
}
Thus, the factor $a$ is just
$a \approx 0.109$ and the asymmetry is therefore suppressed by both $s/m_W^2$ and $a$. Even at
at $\theta = 90^\circ$, where the Born asymmetry is maximal, it is extremely small:
\eq{
    A^{(0)}=\frac{s}{9m_W^2}\frac{a}{s_W^2} \approx 9.4968 \cdot 10^{-8}.
}
We denote the specific contribution to the asymmetry by the index $C$,
which thus can be BSEB, FSEB, VB  or IB=BSEB+FSEB+VB for the whole set of diagrams
Fig.~\ref{fig.Boxeswithvertices}, respectively.

The contribution to the asymmetry $(\Delta A)_C$
and the relative correction $D_A^C$ are defined as:
\eq{
(\Delta A)_C&=\frac{|\M_C^{----}|^2-|\M_C^{++++}|^2}{\sum |\M_0^{\lambda}|^2},
\\
D_A^C &= \frac{(\Delta A)_C}{A^{(0)}} = \frac{|\M_C^{----}|^2-|\M_C^{++++}|^2}{|\M_0^{----}|^2-|\M_0^{++++}|^2}.
}
The relative correction {\bf to observable  asymmetry} from the contribution of type $C$
looks as (see derivation in more details in \cite{Aleksejevs:2012zz}):
\begin{eqnarray}
\delta_A^{C}
= \frac{A_{}^{C}-A_{}^{(0)}}{A_{}^{(0)}}
% =\frac{ \frac{(\sigma^0+\sigma^{C})|_{L-R}}{\sigma^0_{00}+\sigma^{C}_{00}}
%            - \frac{\sigma^0|_{L-R}}{\sigma^0_{00}} }
%      { \frac{\sigma^0|_{L-R}}{\sigma^0_{00}} }
= \frac{ D_A^C -\delta^C }{1+\delta^C},
\label{addit}
\end{eqnarray}
where
%% operation $\sigma|_{L-R}$ means  $\sigma_{LL}-\sigma_{RR}$ and
the relative correction to unpolarized  cross section
$\sigma^0_{00}$ (we used short notation for differential cross section $\sigma \equiv d\sigma/d(\cos\theta)$)
is:
\eq{
    \delta^C= \frac{\sigma^{C}_{00}}{\sigma^0_{00}}.
}
For the two-loop effects where  $\delta^C$ is small, we can use
an approximate equation for relative correction to asymmetry $\delta^{C}_A  \approx  D_A^C$.

% =========================================================================
\section{Insertion of mass operator, vertex and
 vacuum polarization functions to the box type ampli\-tu\-de}
 \label{sec.B}
% =========================================================================

The numerical value of loop momentum squared $|k^2|$ in the box-type ampli\-tu\-des
with the heavy boson exchange  is large compared with the square of
electron mass $|k^2| \gg m^2$, since if $|k^2|$ is far from $M_{Z,W}^2$ the contribution is suppressed with the mass
of heavy boson squared in denominator.
So we can use the asymptotic expressions
for the one-loop vertex functions as well as the mass
and vacuum polarization operators.
Using the well-known approach
%\cite{Berends:NPB206-1982-53},
\cite{Berends:1981uq}, \cite{Berends:1981rb}
which we successfully employed for the box-type chiral amplitudes in \cite{Aleksejevs:2012sq}
(see also \cite{Ahmadov:2012se}),
we can write for the direct $ZZ$-box chiral amplitude of ``++++'' type:
\eq{
\bar{u}_3\gamma_\mu (a+\gamma_5) \dd{k}\gamma_\nu (a+\gamma_5) u_1 \ \bar{u}_4\gamma^\mu
(a+\gamma_5) (-\dd{k})\gamma^\nu (a+\gamma_5) u_2& =
\nn \\
=
-\frac{(a + 1)^4}{g f} \Tr\brs{\dd{p}_3\gamma_\mu\dd{k}\gamma_\nu\dd{p_1}\dd{p}_2\dd{p}_4
\gamma^\mu\dd{k}\gamma^\nu\dd{p}_2\dd{p}_1\omega_+}
=-\frac{8k^2s^2t (a + 1)^4}{g f}.
\label{sa-box}
%
% \bar{u}_3\gamma_\mu\dd{k}\gamma_\nu u_1 \ \bar{u}_4\gamma^\nu\dd{k}\gamma^\mu u_2&=
% \frac{1}{g f} \Tr\brs{\dd{p}_3\gamma_\mu\dd{k}\gamma_\nu\dd{p_1}\dd{p}_2\dd{p}_4
% \gamma^\nu\dd{k}\gamma^\mu\dd{p}_2\dd{p}_1\omega_+}=\frac{2k^2s^2t}{g f}.
}
Easily we can get a similar expression for the crossed box and for amplitude of ``$-\!-\!- -$'' type.
The  quantities $g$ and $f$ in (\ref{sa-box}) coincide with $a$ and $b$ from \cite{Aleksejevs:2012sq},
respectively, and are defined as:
\eq{
g = \bar u_1 \omega_- \dd{p}_2 \omega_+ u_4, \qquad
f = \bar u_2 \omega_- \dd{p}_1 \omega_+ u_3, \nn
}
where $\omega_\pm = \br{1 \pm \gamma_5}/2$ are the chirality projection operators.
Let us calculate $t$-channel amplitude; the $u$-channel amplitude can be obtained
by the replace\-ment $t \leftrightarrow u$.
This interchange will be denoted below as an operator $P_{tu}$.

The box-type amplitude with the double $Z$-boson exchange with all
the possible insertions (i.e. VB, FSEB and BSEB) has a form:
\eq{
\M^{ZZ,\pm}=i\frac{\alpha^2(1 \pm a)^4}{(4c_Ws_W)^4}\frac{6s^2t}{g f m_Z^2}
\int\limits_0^\infty\frac{d\tau}{(1+\tau)^2}I_{ZZ}(\tau),
\label{eq.ZZ}
}
where ``$\pm$'' sign corresponds to the chiral amplitudes $\M^{\pm\pm\pm\pm}$.
The expression for the box amplitude with $Z\gamma$-exchange is similar:
\eq{
\M^{Z\gamma,\pm}=i\frac{2\alpha^2(1 \pm a)^2}{(4c_Ws_W)^2}\frac{6s^2t}{g f m_Z^2}
\int\limits_0^\infty\frac{d\tau}{\tau(1+\tau)}I_{Z\gamma}(\tau).
\label{eq.gZ}
}
At last, for $\gamma\gamma$-exchange amplitude we have:
\eq{
\M^{\gamma\gamma,\pm}=i {\alpha^2}\frac{6s^2t}{g f {m_Z^2}}
\int\limits_z^\infty\frac{d\tau}{\tau^2}I_{\gamma\gamma}(\tau).
\label{eq.gg}
}
In all the above cases, the integration variable is related to the loop momentum as $\tau=-k^2/m_Z^2$.
The lower limit of integration $z=-t/m_Z^2$ for $\M^{\gamma\gamma}$ is introduced
to avoid the double counting for the region of small loop momenta
squares $-k^2 < s$, where we use the Yennie--Frautchi--Suura approach \cite{Yennie:1961ad}.
Finally, the contribution to $\M^{----}$ arises from the box-type Feynman dia\-gram with two $W$-boson exchange:
\eq{
\M_{WW}^{----}=-i\frac{\alpha^2}{2s_W^4}\frac{s^2t}{g f m_W^2}
\int\limits_0^\infty \frac{d\tau_W}{(1+\tau_W)^2}I_{WW}(\tau_W),\qquad \tau_W=-k^2/m_W^2.
\label{eq.WW}
}

The structure of the quantities $I_{ij}$ in (\ref{eq.ZZ}), (\ref{eq.gZ}), (\ref{eq.gg}) and (\ref{eq.WW})
corresponds to three types of radiative corrections, FSEB, VB and BSEB, respectively:
\eq{
I_{ZZ}&=2\M_e + 4V_{eeZ} +2\Pi_{ZZ}, \nn \\
I_{Z\gamma}&=2\M_e+2V_{eeZ}+2V_{ee\gamma}+\Pi_{ZZ}+\Pi_{\gamma\gamma}, \label{eq.Iij} \\
I_{\gamma\gamma}&=2\M_e+4V_{ee\gamma}+2\Pi_{\gamma\gamma}, \nn \\
I_{WW}&=2\M_\nu+4V_{e\nu W}+2\Pi_{WW}. \nn
}
Here, we use the dimensionless quantities for the product of fermion Green function and the truncated mass operators
of electron $M_e$ and neutrino $M_\nu$ (see Appendix~\ref{appendix.FSEB}):
\eq{
\M_{e,\nu}=\frac{i\dd{k}}{k^2} M_{e,\nu}.
}
The vertex function $V^\mu_{ee\gamma}(k^2)$ with one electron on the mass shell and another electron off the mass shell is normalized as
\eq{
V^\mu_{ee\gamma}(k^2)=-ie\gamma^\mu \ V_{ee\gamma}(k^2), \qquad V_{ee\gamma}(0)=0.
}
The vertex function $V^\mu_{eeZ}(k^2)$ is normalized at the point $k^2=m_Z^2$:
\eq{
V^\mu_{eeZ}(k^2)=-\frac{i e}{4c_W s_W}\gamma^\mu \ V_{eeZ}(k^2), \qquad V_{e e Z}(m_Z^2)=0,
}
and similarly for $e \nu W$-vertex function we have:
\eq{
V^\mu_{e\nu W}(k^2)=\frac{i e}{\sqrt{2}s_W}\gamma^\mu\omega_-V_{e\nu W}(k^2), \qquad V_{e\nu W}(m_W^2)=0.
}
% Possible contributions to vertex function which is proportional to 4-vector of photon
% (or $Z$ and $W$ bosons) give zero contribution due to
% gauge invariance of the total amplitude.
The explicit expressions for the vertices $V_{ee\gamma}$, $V_{eeZ}$ and $V_{e\nu W}$ are given in Appendix~\ref{appendix.VB}.

The dimensionless products of boson Green function with the relevant regularized polarization operator $\Pi_{\mu\nu}(q)=\Pi(q^2)g_{\mu\nu}+B(q^2)q_\mu q_\nu$ are defined as:
\eq{
\Pi_\gamma&=\frac{-i}{q^2}\Pi^{\rm tr}_{\gamma\gamma}(q^2), &\qquad \Pi^{\rm tr}_{\gamma\gamma}(0)&
=\frac{\partial}{\partial q^2} \Pi^{\rm tr}_{\gamma\gamma}(0)=0; \nn \\
\Pi_Z&=\frac{-i}{q^2-m_Z^2}\Pi^{\rm tr}_{ZZ}(q^2), &\qquad \Pi^{\rm tr}_{ZZ}(m_Z^2)&=\frac{\partial}{\partial q^2}\Pi^{\rm tr}_{ZZ}(m_Z^2)=0; \nn \\
\Pi_{Z\gamma}&=\frac{-i}{q^2}\Pi^{\rm tr}_{Z\gamma}(q^2), &\qquad \Pi^{\rm tr}_{Z\gamma}(0)&=0; \nn \\
\Pi_W&=\frac{-i}{q^2-m_W^2}\Pi^{\rm tr}_{WW}(q^2), &\qquad \Pi^{\rm tr}_{WW}(m_W^2)&=\frac{\partial}{\partial q^2}\Pi^{\rm tr}_{WW}(m_W^2)=0.
}
The structure $B(q^2) q_\mu q_\nu$ does not contribute due the gauge invariance.
The explicit expression for the ``truncated'' quantities are given in Appendix~\ref{appendix.BSEB}.

% =========================================================================
\section{Numerical results and conclusion}
\label{sec.NumbersAndConclusion}
% =========================================================================

For the numerical calculations, we use the central kinematical point of the MOLLER experiment
and $\alpha$, $m_W$ and $m_Z$ in accordance with the Particle Data Group \cite{Amsler:2008zzb}.
The effective quark masses used for the vector boson self-energy loop contributions
are extracted from the shifts in the fine structure constant due to hadronic vacuum polarization
$\Delta \alpha_{\rm had}^{(5)}\br{m_Z^2} = 0.02757$ \cite{Jegerlehner:2001wq}.
For the mass of Higgs boson, we take $m_H = 125~\GeV$.

The contribution relevant to the observed asymmetry is the interference of the
two-loop box-type amplitudes with the Born amplitudes $\M_{\gamma,Z}$.
The contribution to the matrix element squared (i.e. cross section) has the form:
\eq{
|\M^{\pm\pm\pm\pm}_{\rm IB}|^2&
=2\br{1+P_{tu}}\brs{\br{\M^{ZZ}+\M^{Z\gamma}+\M^{WW}}\M_\gamma^*+\M^{\gamma\gamma}\M_Z^*}.
}
In the right-hand side of this equation, we assume that the amplitudes are taken
in the same chiral state corresponding to the state of left-hand side.
Note that the intermediate states with $W^{\pm}$ bosons and Faddeev--Popov ghosts $G_W^{\pm}$ contribute
to the mass and vertex operators in the $\M^{----}$ chiral amplitude.
Since the parameter $a$ is very small, we can present the final result as:
\eq{
\brm{\M^{----}_{\rm IB}}^2 - \brm{\M^{++++}_{\rm IB}}^2
=-H(a)+\br{H(-a)+Y} = - 2a\Bigl.\frac{\partial H(a)}{\partial a}\Bigr|_{a\rightarrow0} + Y,
}
and thus the relative correction $D_A^{\rm IB}$ has the form:
\eq{
D_A^{\rm IB} & = \frac{t^2u^2}{128\br{\pi\alpha}^2
\br{s^4+t^4+u^4}}\Bigl( - 2a\Bigl.\frac{\partial H(a)}{\partial a}\Bigr|_{a\rightarrow0} + Y \Bigr)
\frac{1}{A^{(0)}}.
\label{eq.DA}
}
We define $H$ and $Y$ as:
\eq{
H&=H_{ZZ}+H_{Z\gamma}+H_{\gamma\gamma}+H_{WW}+H_{\rm mix},
\\
Y&=Y_{ZZ}+Y_{Z\gamma}+Y_{\gamma\gamma}+Y_{WW}+Y_{\rm mix},
\nn
}
where the first four terms in both $H$ and $Y$ correspond to the box-type amplitudes
with $ZZ$, $Z\gamma$, $\gamma\gamma$ and $WW$ bosons exchanged between electrons, and
the last term corresponds to the cases with $Z$ or $\gamma$ and the mixed boson
Green function with polarization operator $\Pi_{Z\gamma}$.

Using the following relations (see, for example, \cite{Aleksejevs:2012sq} and \cite{Ahmadov:2012se})
\eq{
\frac{1}{g f}\br{\frac{1}{g f}-\frac{1}{c d}}^*
=-\frac{1}{st^2u};\qquad
\frac{1}{g f}\br{\frac{t}{g f}-\frac{u}{c d}}^*=\frac{2}{s^2t}
}
we obtain the following numerical results:
\eq{
H_{ZZ}&=-\frac{3\alpha^3\pi(1+a)^4}{8(c_Ws_W)^4}\frac{s^3}{m_Z^2tu}(1+P_{tu})\int\limits_0^\infty\frac{d\tau}{(1+\tau)^2}
\left[2(M_e^\gamma+(1+a)^2M_e^Z) + \right.\nn \\
&\qquad\left. +2\Pi_{ZZ}+4(V^\gamma+(1+a)^2V^Z)\right]
=
\nn\\
&=
-1.653 \cdot 10^{-13}\br{1 + a}^4 \br{-81.36 - 1.1293 \br{1 + a}^2} \nn \\
Y_{ZZ}&=-\frac{3\alpha^3\pi}{8(c_Ws_W)^4}\frac{s^3}{m_Z^2tu}(1+P_{tu})\int\limits_0^\infty\frac{d\tau}{(1+\tau)^2}[2M_e^W+4(V^\nu+V^{2\nu})] =
\nn\\
&= 3.139 \cdot 10^{-12};       \nn \\
H_{Z\gamma}&=-\frac{12\alpha^3\pi(1+a)^2}{(c_Ws_W)^2}\frac{s^3}{m_Z^2tu}(1+P_{tu})\int\limits_0^\infty
\frac{d\tau}{\tau(1+\tau)}[2(M_e^\gamma+(1+a)^2M_e^Z) + \nn \\
&\qquad+\Pi_{ZZ}+\Pi_{\gamma\gamma}+2(V^\gamma+V_{ee\gamma}^\gamma+(1+a)^2(V^Z+V_{ee\gamma}^Z))]
=
\nn\\
&=
-9.155 \cdot 10^{-11} \br{1 + a}^2 \br{ -4.30744 - 0.04567 \br{1 + a}^2};    \nn \\
Y_{Z\gamma}&=-\frac{12\alpha^3\pi}{(c_Ws_W)^2}\frac{s^3}{m_Z^2tu}(1+P_{tu})\int\limits_0^\infty\frac{d\tau}{\tau(1+\tau)}[2M_e^W+2(V^\nu+V^{2\nu}+V_{ee\gamma}^W)]
=
\nn\\
&=
6.974 \cdot 10^{-11}, \nn
}
\eq{
H_{\gamma\gamma}&=\frac{12\alpha^3\pi(1+a)^2}{(c_Ws_W)^2}(1+P_{tu})\frac{s^2 z}{m_Z^2t}\int\limits_z^\infty\frac{d\tau}{\tau^2}[2(M_e^\gamma+(1+a)^2M_e^Z) + \nn \\
&\qquad+2\Pi_{\gamma\gamma}+4(V_{ee\gamma}^\gamma+(1+a)^2V_{ee\gamma}^Z)]
=
\nn\\
&=
-3.094\cdot 10^{-12} (1 + a)^2 (-2.52038 - 5.04456 \cdot 10^{-6} (1 + a)^2),     \nn \\
Y_{\gamma\gamma}&=\frac{12\alpha^3\pi}{(c_Ws_W)^2}(1+P_{tu})\frac{s^2 z}{m_Z^2t}\int\limits_z^\infty\frac{d\tau}{\tau^2}[2M_e^W+4V_{ee\gamma}^W]
=-4.4261 \cdot 10^{-17}, \nn  \\
H_{WW}&=0; \nn \\
Y_{WW}&=\frac{8\alpha^3\pi}{(s_W)^4}\frac{s^3}{m_Z^2tu}(1+P_{tu})\int\limits_0^\infty\frac{d\tau_W}{(1+\tau_W)^2}[2M_\nu+2\Pi_{WW}+4V_{eW\nu}]
=
\nn\\
&=
-3.36\cdot 10^{-10}.
}

The ``mixed''-type amplitude in two-loop approximation has two different contributions $(H,Y)_{\rm mix}=(H,Y)^{(1)}_{\rm mix}+(H,Y)^{(2)}_{\rm mix}$.
The first contribution is associated with the two-loop box-type amplitude:
\eq{
H^{(1)}_{\rm mix}&=-\frac{6\alpha^3\pi(1+a)}{(c_Ws_W)^3}\frac{s^3}{m_Z^2tu}(1+P_{tu})\int\limits_0^\infty
\frac{d\tau_W}{(1+\tau_W)}R_{\gamma Z}(\tau_W)\times
\nn\\
&\times
\br{(4c_Ws_W)^2\frac{1}{\tau_W}+(1+a)^2\frac{1}{1+\tau_W}}= \nn\\
&=-1.10029 \cdot 10^{-9} (1 + a) (0.007746- 0.000340 (1 + a)^2), \nn \\
Y^{(1)}_{\rm mix}&=0, \nn \\
R_{\gamma Z}(\tau_W)&=\frac{\alpha c_W}{8\pi s_W}\br{-3+2(3-2c_W^2)\frac{1}{\tau_W}}L_W(\tau_W), \nn\\
L_W(\tau_W)&=\int\limits_0^1d x\log(1+x(1-x)\tau_W). \nn
}
The second contribution arises from the interference of the Born-type amplitude with the mixed Green function
and the box type one-loop amplitude with the $\gamma\gamma$-exchange:
\eq{
H^{(2)}_{\rm mix}&=\frac{48\alpha^3\pi(1+a)}{c_Ws_W}(1+P_{tu})\frac{s^3 z}{t^2u} R_{\gamma Z}\br{z}
=
-3.982 \cdot 10^{-13} (1 + a), \nn\\
Y^{(2)}_{\rm mix}&=0.
}

The contributions to the asymmetry from the transition polarization operator $\Pi_{Z\gamma}$
with leptons in the fermion loop are proportional to higher
powers of $a$, which is small.
The same reasoning is valid for the quark-antiquark state contribution. Specifically, it enters with the factor
\eq{
(2/3)(1-(8/3)s_W^2)-(1/3)(1-(4/3)s_W^2)=a/3.
}
The contributions from $(W^+W^-),\ (W^\pm G_W^\mp),\ (G_W^\pm G_W^\mp)$ intermediate states are
considered in Appendix~\ref{appendix.BSEB}.

%Using the expression for the truncated transition polarization
%operator given in Appendix~\ref{appendix.D} we put the contribution to the asymmetry
%We obtain for the correction factor the following result
%\eq{
%D_A^{\rm IB} = -0.000526.
%}

Finally, we are ready to present final numerical value for the relative corrections considered in this paper to the observable cross section asymmetry.
The one-loop (NLO) corrections \cite{Aleksejevs:2010ub,Aleksejevs:2012zz} give the biggest contribution,
\eq{
    \delta_A^{\text{NLO}} = -0.6953.
}
Several categories of the NNLO contributions ($Q$-part and double boxes) are calculated
in \cite{Aleksejevs:2011de} and \cite{Aleksejevs:2012sq} and give the following values:
\eq{
    \delta_A^{\rm NLO+Q} = -0.6535,
    \qquad
    \delta_A^{\text{double box}} \approx D_A^{\text{double box}} = -0.0101.
}
Summing up all the contributions in (\ref{eq.DA}), the numerical result of the class of the gauge-invariant Feynman amplitudes considered in this paper (boxes with one-loop insertions of
fermion mass operators,
vertex functions and polarization of vacuum for bosons) is:
\eq{
\delta_A^{\text{IB}} \approx D_A^{\rm IB} = -0.0039.
%1.79633 \cdot 10^{-9},
%%  + 2.0496 \cdot 10^{-12} \, C_{\rm BMR},
}
%that is
%\eq{
%\delta_A^{\rm this\ paper}
%=\frac{(\Delta A)_I+(\Delta A)_{II}}{A^{(0)}}=
%0.00431869,
%%0.0189.
%%% + 2.1582 \cdot 10^{-5} \, C_{\rm BMR},
%}
%The expected MOLLER experimental error is about two per cents,
%so we can see, the size of the obtained relative correction is
%much less (by factor $\sim 5$) but still important as contribution to the MOLLER error budget.
As one can see, the relative correction we obtained is
much less than the expected MOLLER experimental error, but it still a non-negligible contri\-bu\-tion to the MOLLER error budget. Most likely, the entire set of two-loop corrections will be smaller than the experimental statistical error, but, in the light of the MOLLER success depending so crucially on its precision, the two-loop corrections still need to be controlled.

As the low-energy precision experiment, MOLLER is complementary to the LHC efforts and may discover
new physics signal that could escape LHC detection.
However, for the MOLLER experiment to produce meaningful physics, the uncertainties in the NNLO EWC must be much smaller than the MOLLER
statistical error. Clearly, there is a need
for the a complete study of the two-loop electroweak radiative corrections in order
to meet the MOLLER precision goals.

% =========================================================================
\section{Acknowledgements}
% =========================================================================
Many thanks to A.~I.~Ahmadov and D.~Yu.~Bardin for help and valuable discussions.
This work is supported by the Natural Sciences and Engineering Research Council of Canada,
Harrison McCain Foundation Award,
Belarus scien\-tific program ``Convergence'', and RFBR grant 11-02-00112.
VAZ is grate\-ful to Acadia, Memorial Universities, and JINR for hospitality.

\appendix

% =========================================================================
\section{Mass operators}
\label{appendix.FSEB}
% =========================================================================

Here, we will define the explicit form of the quantities $M_e$, $M_\nu$
which enter to $I_{ij}$ from (\ref{eq.Iij}). The quantity $M_e$ has the following form:
\eq{
M_e=M_e^\gamma+(a \pm \gamma_5)^2 M_e^Z+ \omega_-^2 M_e^W.
}
The explicit expression for the truncated mass operator in QED was found by
R.~Karplus and N.~Kroll in 1950 \cite{Karplus:1950zzb,Akhiezer:1981}:
\eq{
M_e^\gamma&=\frac{i \alpha}{2\pi m}\br{\dd{p}-m}^2
\left[
    \frac{1}{2(1-\rho)}
    \br{1-\frac{2-3\rho}{1-\rho}\log \rho}-
\right.
\nn\\
&-\left.
    \frac{\dd{p}+m}{m \rho}
    \br{\frac{1}{2(1-\rho)}
        \br{2-\rho + \frac{\rho^2+4\rho-4}{1-\rho}\log \rho}+1+2\log\frac{\lambda}{m}
    }\right],
\\
    \rho&=1-\frac{p^2}{m^2}. \nn
}
It is useful to note that the expression in the square brackets is finite at $\rho \to 1$.
In the limit of large $\tau_1=-p^2/m^2$ with logarithmical accuracy we have
\eq{
    M_e^\gamma =  M_e^\gamma(\tau_1)\cdot \dd{p} \approx -\frac{\alpha}{4\pi}\log\br{\tau_1} \cdot \dd{p},
    \qquad
    \tau_1 \gg 1.
}
%\eq{
%M_e^\gamma = \dd{p} M_e^\gamma(\tau_e),\qquad
%M_e^\gamma(\tau_e)=-\frac{\alpha \tau_e}{4\pi(1+\tau_e)}\log \tau_e.
%}
This mass operator contribution to the integral in (\ref{eq.gg}) with logarithmical accuracy gives:
\eq{
-\frac{t}{m_Z^2}\int\limits_{-t/m_Z^2}^\infty\frac{d \tau}{\tau^2}M_e^\gamma(\tau_1)
 = - \frac{\alpha}{4\pi}  \log\frac{-t}{m^2}.
}
The mass operators induced by additional $Z$ and $W$ bosons have the following form:
\eq{
M_e^Z&=\frac{\alpha}{2\pi(4c_Ws_W)^2}\int\limits_0^1 (1-x)\log(1+\tau x) d x,  \nn \\
M_\nu&=M_e^W=\frac{\alpha}{\pi s_W^2}\int\limits_0^1 (1-x)\log(1+\tau x) d x.
}

% =========================================================================
\section{Vertices}
\label{appendix.VB}
% =========================================================================

The general form of the vertex function is $V^\mu(k)=A\gamma^\mu+B k^\mu$;
the term $B k^\mu$ inserted in the box-type amplitude gives no contribution
due to the gauge invariance.
The vertex function with one electron on the mass shell and other electron
off the mass shell $V_{ee\gamma}^\mu(p, p-k, k)=-ie\gamma^\mu V_{ee\gamma}(k^2)$, normalized as $V_{ee\gamma}(0)=0$,
has three contributions:
\eq{
    V_{ee\gamma}=V_{ee\gamma}^\gamma + (a\pm \gamma_5)^2 V_{ee\gamma}^Z + \omega_-^2 V_{ee\gamma}^W.
}
First, let us consider the QED-type contribution with the virtual photon inter\-mediate
state $V_{ee\gamma}^\gamma$.
The standard procedure of joining
denominators and per\-for\-ming the loop momenta integration leads to
\eq{
V_{ee\gamma}^\gamma(k^2)&=\frac{\alpha}{4\pi}\int\limits_0^1 d x\int\limits_0^1 y d y \br{\log\frac{\Lambda^2}{D}+\frac{k^2 b\bar{b}}{2D}},
\qquad
b=x y, \qquad \bar{b}=1-b,\\
D&=(m^2-k^2x(1-x))y^2+(1-y)\lambda^2-y(1-y)(k^2-2p_1k), \nn
}
where $\Lambda$ is cut-off regularization parameter.
%We applied here this cut-off technique instead of probably more popular dimensional regularization scheme
%(which in addition provides gauge invariance).
%Difference between both schemes is numerically nonsignificant.
Since the sub-set of the diagrams considered here is gauge invariant on its own, it was not essential for us to use the dimensional regularization
scheme providing gauge invariance, so we simply applied the cut-off technique. There is no significant numerical difference between two schemes in this situation.

The renormalization procedure consists in subtraction at $k=0$ and leads to:
\eq{
V_{ee\gamma}^\gamma\br{\tau_e}&
= - \frac{\alpha}{4\pi}\int\limits_0^1 \log({1+x(1-x)\tau_e}) d x, \qquad \tau_e=-\frac{k^2}{m^2}.
}
The contribution of this vertex function to the integral in (\ref{eq.gg}) has the form:
\eq{
-\frac{t}{m_Z^2}\int\limits_{-t/m_Z^2}^\infty\frac{d \tau}{\tau^2}V_{ee\gamma}^\gamma(\tau_e)
\approx \frac{\alpha}{4\pi}\br{1-\log\frac{-t}{m^2}}.
}
The other contributions are:
\eq{
V^Z_{ee\gamma}&=\frac{\alpha}{2\pi(4c_Ws_W)^2}\int\limits_0^1d x\int\limits_0^1y d y \br{\log\frac{1-y}{1-y+b\bar{b}\tau}-\frac{b\bar{b}\tau}{2(1-y+b\bar{b}\tau)}},\nn \\
V^W_{ee\gamma}&=\frac{\alpha}{4\pi s_W^2}\int\limits_0^1d x\int\limits_0^1y d y \br{3\log\frac{y c_W^2+\tau b\bar{b}}{y c_W^2}-\frac{\tau b(\bar b - b)}{2(y c_W^2+\tau b\bar{b})}}.
\label{eq.Veeg}
}
Vertex function $V^\mu_{eeZ}=-iG\gamma^\mu(a\pm \gamma_5)V_{eeZ}$, $G=e/(4s_Wc_W)$
has four different contributions:
\eq{
    V_{eeZ}=\omega_-V^\gamma+(a\pm \gamma_5)^2V^Z+\omega_-V^\nu+V^{2\nu},
}
and is normalized as $V_{eeZ}(k^2=m_Z^2)=0$.
These contributions are
\eq{
V^\gamma&=-\frac{\alpha}{4\pi}\log\tau,\nn \\
V^Z&=\frac{1}{(4c_Ws_W)^2}\frac{\alpha}{2\pi}\int\limits_0^1d x\int\limits_0^1y dy \times\nn\\
&\times\br{\log\frac{1-y-b\bar{b}}{1-y+b\bar{b}\tau}-\frac{b\bar{b}\tau}{2(1-y+b\bar{b}\tau)}-\frac{b\bar{b}}{2(1-y-b\bar{b})}}, \\
V^\nu&=-\frac{\alpha c_W^2}{4\pi}\int\limits_0^1d x\int\limits_0^1y d y \times\nn\\
&\times\br{3\log\frac{yc_W^2+\tau b\bar{b}}{y c_W^2-b\bar{b}}-\frac{\tau b(\bar b - b)}{2(y c_W^2+\tau b\bar{b})}-\frac{b(\bar b - b)}{2(y c_W^2-b\bar{b})}}, \nn \\
V^{2\nu}&=\frac{\alpha}{2\pi s_W^2}\int\limits_0^1d x\int\limits_0^1y d y \times\nn\\
&\times\br{\log\frac{(1-y)c_W^2- b\bar{b}}{(1-y)c_W^2+\tau b\bar{b}}-
\frac{\tau b\bar{b}}{2(y c_W^2+\tau b\bar{b})}-\frac{b\bar{b}}{2(y c_W^2- b\bar{b})}}.\nn
}

And finally, the vertex function $V^\mu_{e \nu W}=i\frac{\gamma_\mu\omega_-}{\sqrt{2}} V_{e \nu W}$
as well contains three contributions:
\eq{
    V_{e\nu W} =V^{ZW}+V^{WZ}+V^{\gamma W},
}
and is normalized as $V_{e\nu W}(\tau=-c_W^2)=0$ and $V^{ZW}=V^{WZ}$.
So the contribu\-tions are:
\eq{
V^{ZW}&=\frac{\alpha}{4\pi s_W^2}\int\limits_0^1d x\int\limits_0^1 y d y \times\nn\\
&\times\left(-3\log\frac{y a_x+\tau b\bar{b}}{y a_x-c_W^2b\bar{b}}+
+\frac{\tau b(\bar b - b)}{2(y a_x+\tau b \bar b)}+\frac{c_W^2b(\bar b - b)}{2(y a_x-c_W^2b \bar b)}\right), \nn \\
V^{\gamma W}&=-\frac{\alpha}{4\pi}\left[\int\limits_0^1d x \int\limits_0^1d y \right.\times\nn\\
&\times\left.\left(3\log\frac{b c_W^2+\tau+c_W^2}{b c_W^2}+ \frac{\tau (\bar b - b)}{2(c_W^2+\tau \bar b)}\right)
-1+\frac{1}{4}\log\frac{m_W^2}{m^2}+\frac{1}{4}\log\frac{m^2}{\lambda^2}\right],\nn
}
where $a_x=x+(1-x)c_W^2$.
Note that the term containing $\log(m^2/\lambda^2)$ in expression $V^{\gamma W}$ can be omitted as it will be absorbed by the
similar terms in two-loop contributions after applying the Yennie--Frautschi--Suura regulari\-za\-tion (see \cite{Aleksejevs:2012xua} for details).

% =========================================================================
 \section{Polarization operators}
\label{appendix.BSEB}
% =========================================================================

While considering the vacuum polarization operators of photon, $Z$- and $W$-boson at one loop,
one should recall that the regularization implies the double subtraction procedure.
The ``truncated'' operators imply including only the ver\-tices of interaction of bosons with the fermion loop.
From now on, we will omit index ``tr''.
The general form of the polarization operator is:
\eq{
    \Pi_{\mu\nu}(q)=g_{\mu\nu}\Pi(q^2)+q_\mu q_\nu B(q^2).
}
We only need to consider a part of polarization tensor proportional to $g_{\mu\nu}$.
The reason is the gauge invariance of the whole set of the double-box ampli\-tu\-des, which leads to a zero contribution for terms proportional to $q_\mu q_\nu$ tensor.

Let's define $\Pi^\gamma$ as:
\eq{
    \Pi^\gamma=-\frac{i}{q^2}\Pi^\gamma(q^2).
}
It has five types of contributions, corresponding to the intermediate state of lepton--antilepton pairs,
quark--antiquark pairs, $W^+W^-$ and the charged ghost state $G^+_WG^-_W$:
\eq{
\Pi^\gamma=\Pi^l+\Pi^q+\Pi^{WW}+\Pi^{G_W^\pm G_W^\mp}+\Pi^{W^\pm G_W^\mp}.
\label{eq.PiGamma}
}
The contribution of leptons and quarks are associated with the quadratic
di\-ver\-gent integral over the loop momentum:
\eq{
\frac{1}{4} \int  \frac{d k}{\br{k^2-m^2}\br{(k-q)^2-m^2}} \Sp\brs{(\dd{k}+m)\gamma_\mu(\dd{k}-\dd{q}+m)\gamma_\nu}.
}
Using the set of divergent integrals (see Appendix~\ref{appendix.Integrals})
and performing the re\-gu\-la\-ri\-za\-tion procedure, we include the contribution of leptons and quarks as
\eq{
\Pi^l+\Pi^q=\frac{\alpha}{\pi\tau}\br{\sum_{l=e,\mu,\tau}G(\tau, \sigma^l)+3\sum_{q=u,d,s,\cdots}Q_q^2G(\tau,\sigma^q)},
}
where
\ga{
G(\tau,\sigma)=
\frac{1}{3}(\tau-2\sigma)L(\frac{\tau}{\sigma})+\frac{1}{9}\tau,\qquad L(z)=\int\limits_0^1d x \log(1+x(1-x)z), \nn\\
\sigma^f=\frac{m_f^2}{m_Z^2},\qquad \tau=-\frac{q^2}{m_Z^2}, \nn
}
%It is a real quantity and we introduce the general factor
%$-i/q^2$ to provide insertion polarization operator to the box amplitude.
The factor 3 takes into account the number of quark colours.
The last three contributions in (\ref{eq.PiGamma}) are
\eq{
\Pi^{WW}+\Pi^{G_W^\pm G_W^\mp}+\Pi^{W^\pm G_W^\mp}=-\frac{\alpha}{12\pi \tau}\br{\frac{1}{6}\tau+(5\tau-c_W^2)L\br{\frac{\tau}{c_W^2}}},
}
the known result for Feynman--t'Hooft gauge used in \cite{Cheng1988,Denner:1991kt}.
%(Please note that the Feynman rule for charged ghosts-photon vertex in  \cite{Cheng1988} has an typo; see
%the correct expression in A.17 of \cite{Denner:1991kt}.)

The polarization operator for $Z$-boson has seven types of contributions:
\eq{
\Pi^Z=\Pi_Z^l + \Pi_Z^q + \Pi_Z^\nu + \Pi_Z^{W^+W^-} + \Pi_Z^{G_W^+G_W^-} + \Pi_Z^{G_1 G_2} + \Pi_Z^{W^\pm G_W^\mp},
}
were we used the definition
\eq{
\Pi^Z=-\frac{i}{q^2-m_Z^2}\Pi^Z_{\rm tr}(q^2).
}
The contribution of lepton $\Pi_Z^l$, quark $\Pi_Z^q$ and the neutrino $\Pi_Z^\nu$ loops
%(here and further we take into account the form of box amplitude and the two boson propagators attached to the truncated
%polarization operator $\Pi(q^2)$ which results in factor $(-i/(q^2-m_Z^2))\Pi(q^2)=\Pi_{ZZ}(q^2)$)
can be calculated in the non-renormalized  approach:
%For electron loop we obtain for un re-normalized contribution
\eq{
\Pi_{\mu\nu}=\frac{\alpha}{12\pi}\br{q^2\log\frac{\Lambda^2}{q^2}+O(q^2)}g_{\mu\nu}.
}
The renormalization of $R(\tau)$ for any contribution to the polarization operator of $Z$-boson consist of the following replacement:
\eq{
R(\tau) \to R(\tau)-R(-1)-(\tau+1)R'(-1).
}
In particular, for example:
\eq{
-q^2\log\frac{q^2}{m^2} \to m_Z^2F(\tau), \qquad F(\tau)=\tau\log\tau-1-\tau.
\label{eq.F}
}
Keeping in mind that there are three generations of charged leptons, neutrinos, and quarks,
% which is associated with factor $1/(2s_Wc_W)^2$ and contribution of $u,c$ and $d,s$
%quarks with factors
%$$ \frac{1}{(4s_Wc_W)^2}[1+(1-\frac{8}{3}s_W^2)^2],\frac{1}{(4s_Wc_W)^2}[1+(1-\frac{4}{3}s_W^2)^2], $$
we obtain:
\eq{
\Pi^{l+q+\nu}_Z=\frac{\alpha}{12\pi}\frac{F(\tau)}{1+\tau}\brs{3+\frac{3}{4(s_Wc_W)^2}+\frac{1}{2(s_Wc_W)^2}\br{1-2s_W^2+\frac{20}{9}s_W^4}}.
}
The contribution of $W^+W^-$ pair in the intermediate state to the $Z$-boson polarization operator looks like:
\eq{
\Pi_Z^{W^+W^-}&=\frac{\alpha c_W^2}{8\pi s_W^2}\frac{1}{1+\tau}
\times\nn\\
&\times\brs{\br{\frac{19}{6}\tau-\frac{16}{3}}L\br{\frac{\tau}{c_W^2}}-\br{\frac{19}{6}\tau-\frac{16}{3}}c_1+ \frac{17}{2}(\tau+1)c_2},
 \nn\\
&c_1=L\br{-\frac{1}{c_W^2}} \approx -0.248, \qquad c_2=\int\limits_0^1\frac{x(1-x)}{1-x(1-x)/c_W^2} d x\approx 0.226. \nn
}
The contribution of the charged ghosts $G_W^\pm$ is:
\eq{
\Pi_Z^{G_W^+G_W^-}&=-\frac{\alpha\br{1-2s_W^2}^2}{4\pi(c_Ws_W)^2}\frac{1}{1+\tau}
\times\nn\\
&\times\brs{\br{\frac{1}{12}\tau+\frac{1}{3}c_W^2}\br{L\br{\frac{\tau}{c_W^2}}-c_1}+
\frac{1}{c_W^2}(\tau+1)\br{\frac{1}{12}-\frac{1}{3}c_W^2}c_2}, \nn \\
\Pi_Z^{W^\pm G_W^\mp}&=-\frac{\alpha s_W^4}{2\pi}\frac{1}{1+\tau}\brs{L\br{\frac{\tau}{c_W^2}}-c_1-\frac{1}{c_W^2}c_2(1+\tau)}. \nn
}
And, finally, the contribution from the state with ghosts $G_{1,2}$ is:
\eq{
\Pi_Z^{G_1 G_2}=\frac{\alpha}{4\pi(c_Ws_W)^2}\frac{1}{1+\tau}[\tau(A(\tau)-A(-1))+(\tau+1)A'(-1)],
}
with explicit form of $A(\tau)$ given in Appendix~\ref{appendix.Integrals}.
% with $\gamma=m_H^2/m_Z^2 = 1.879$ at $m_H=125~\GeV$ and the mass of one of ghost state ($\phi_1$) \cite{Cheng1988}

The polarization operator for $W$-boson has contributions from the loop Feynman diagrams with
$(\nu,e)$, $(\bar{d}+\bar{s})(u+c)$, $(W,Z)$, $(W,\gamma)$ and the states with ghosts.
Defining the dimensionless combination:
\eq{
\Pi^W=-\frac{i}{q^2-m_W^2}\Pi^Z_{\rm tr}(q^2)=\Pi^W(\tau_W), \qquad \tau_W=-\frac{q^2}{m_W^2},
}
we write
\eq{
\Pi^W=\Pi_W^{l\bar{\nu_l}}+\sum_q \Pi_W^q+\Pi_W^{WZ}+\Pi_W^{WG_Z}+\Pi_W^{G_W,G_Z}+\Pi_W^{W,\gamma}+\Pi_W^{G_W,\gamma}.
}
From now on, when considering the definite contributions to $\Pi^W$, we imply that $\tau\to\tau_W.$
Let us first consider the contributions from fermions.
For the state with a charged lepton and the corresponding antineutrino we obtain:
\eq{
\sum\Pi_W^{l\bar{\nu_l}}=3\frac{\alpha}{24\pi s_W^2}\frac{1}{1+\tau}F(\tau),
%\ \tau=\tau_W=-\frac{q^2}{m_W^2},
}
with function $F$ given in (\ref{eq.F}). Factor 3 corresponds to the number of lepton generations.
The contribution of quark states is:
\eq{
\sum_{q=u,d,s,c} \Pi_W^q=4\frac{\alpha}{24\pi s_W^2}\frac{1}{1+\tau}F(\tau),
}
where factor 4 corresponds to the number of pairs $(\bar{d}+\bar{s})(u+c)$.
The for the $WZ$ state we have:
\eq{
\Pi_W^{WZ}&=-\frac{\alpha c_W^2}{4s_W^2(1+\tau_W)}[\Psi(\tau_W)-\Psi(-1)-(1+\tau_W)\Psi'(-1)], \\
\Psi(z)&=\br{4z-1-\frac{1}{c_W^2}}\int\limits_0^1\log\br{x+\frac{1-x}{c_W^2}+x(1-x)z} d x-
\nn\\
&-\br{\frac{1}{12}z+\frac{1}{3}}\int\limits_0^1\log(1+x(1-x)z) d x+\nn \\
&+\frac{1}{2}s_W^2\int\limits_0^1 d x\int\limits_0^1y\log\br{y+(1-y)\br{x+\frac{1-x}{c_W^2}}+y(1-y)z} d y, \nn\\
\Psi(-1) &= 0.226, \qquad \Psi'(-1) = -1.26. \nn
}
Now we consider the intermediate states $(W,G_Z)$ and $(G_W,G_Z)$.
For the insertion to the box amplitude we have:
\eq{
&\Pi_W^{WG_Z}+\Pi_W^{G_W,G_Z}=
\nn\\
&\qquad=-\frac{\alpha}{2\pi}\frac{1}{1+\tau}\left[
    \int\limits_0^1 d x\log\br{\frac{x+(1-x)(1+\tau x)c_W^2}{x+(1-x)^2c_W^2}}-
    \right.\nn\\
    &\qquad\qquad-\left.
\tau\br{A(\tau)-A\br{-\frac{1}{c_W^2}}}-(\tau+c_W^2)A'\br{-\frac{1}{c_W^2}}\right],
}
with $A(\tau)$ taken with $\gamma=1/c_W^2$.

For the last two terms we have:
\eq{
\Pi_W^{W,\gamma}+\Pi_W^{G_W,\gamma}&=\frac{\alpha}{4\pi(\tau+1)}\br{-4 Q(\tau)+\frac{5}{36}R(\tau)}, \nn\\
Q(\tau)&=\tau\int\limits_0^1d x\log(1+\tau x)+3+(1+\tau)\br{1+\frac{1}{2}\log\frac{m_Z^2}{\lambda^2}}. \nn\\
R(\tau)&=-\frac{6}{\tau^2}-\frac{15}{\tau}+11+6\br{\frac{1+\tau}{\tau}}^3\log(1+\tau)-20-27(1+\tau), \nn
}
Note that the term $\log\br{m^2/\lambda^2}$ in the expression for $Q(\tau)$ is compensated by the
corresponding contributions from the two-box amplitudes.

Let us now consider the contributions to the transition polarization $\Pi_{\mu\nu}^{Z\gamma}=\Pi^{Z\gamma}g_{\mu\nu}$,
and define the dimensionless function
\eq{
\Pi^{Z\gamma}=-\frac{i}{q^2}\Pi^{Z\gamma}_{\rm tr}(q^2).
}
As shown above, the fermions contribution is proportional to $a^2$ and can be omitted.
The contributions of $(W^+W^-)$, $(W^\pm G_W^\mp)$, $(G_W^\pm G_W^\mp)$ to $\Pi^{Z\gamma}$ are, res\-pec\-ti\-vely:
\ga{
-i\frac{\alpha c_W}{8\pi s_W}\br{-\frac{19}{6}+\frac{16}{3\tau_W}}L(\tau_W),\qquad
-i\frac{\alpha c_W}{8\pi s_W}\br{\frac{1}{6}+\frac{2}{3\tau_W}}L(\tau_W),\nn\\
i\frac{\alpha c^3_W}{2\pi s_W \tau_W}L(\tau_W).\nn
}
Thus, the total is:
\eq{
\Pi^{Z\gamma}=-\frac{i\alpha c_W}{8\pi s_W}\br{-3+\frac{1}{\tau_W}(6-4c_W^2)}L(\tau_W).
}

% =========================================================================
\section{Loop integrals and regularization}
\label{appendix.Integrals}
% =========================================================================

To calculate loop integrals, we perform the Wick rotation of the loop mo\-men\-tum $k$
($k_0\to ik_4$, $k^2=-k_E^2<0$).
In order to regularize ultra-violet divergence, we introduce the cut-off parameter $\Lambda$ so $k_E^2<\Lambda^2$, and all of the kinematical invariants much less (i.e. $\Lambda^2 \gg | p_i p_j |$). The final result will be independent of $\Lambda$ after the renormalization procedure.
Let us now list all the integrals we need:
\eq{
\int\frac{k^2 d k}{(k^2-D)^3}&=\log\frac{\Lambda^2}{D}-\frac{3}{2},&\qquad
\int\frac{d k}{(k^2-D)^2}&=\log\frac{\Lambda^2}{D}-1,\nn \\
\int\frac{d k}{(k^2-D)^3}&=-\frac{1}{2 D}, &\qquad\int\frac{d k}{(k^2-D)^4}&=\frac{1}{6D^2}, \\
\int\frac{(k^2)^2 d k}{(k^2-D)^4}&=\log\frac{\Lambda^2}{D}-\frac{11}{6},
&\qquad\int\frac{k^2 d k }{(k^2-D)^4}&=-\frac{1}{3D}.\nn
}
Here, we use the notation $d k \equiv d^4 k/(i\pi^2)=k_E^2 d k_E^2$, where $k_E$ is the Euclidean 4-vector (i.e. $k_E^2=k_1^2+k_2^2+k_3^2+k_4^2>0$)
and omit the terms of order $O(D/\Lambda^2)$.
We also use the consequence of the integrand symmetry:
\eq{
\int f\br{k^2} k_\mu dk =0
}
for any function $f(k^2)$.
The standard procedure of shifting variable in loop integrals \cite{Akhiezer:1981} leads to:
\eq{
\int\frac{d k}{((k-b)^2-d)^2} = \log\frac{\Lambda^2}{d}-1,
\qquad
\int\frac{k_\mu d k }{((k-b)^2-d)^2} = b_\mu \br{\log\frac{\Lambda^2}{d}-\frac{3}{2}}.
\nn
}
Let us consider the divergent integrals with $A\equiv k^2-m^2$ and $B\equiv (q-k)^2-m^2$:
 \eq{
 \int\frac{d k}{A B}&=L_\Lambda-1-L,\nn\\
 \int\frac{k_\mu d k }{A B} &=\frac{1}{2}q_\mu\br{L_\Lambda-\frac{3}{2}-L}, \nn\\
 \int\frac{k_\mu k_\nu d k }{A B}&=g_{\mu\nu}\left\{-\frac{\Lambda^2}{4}+\frac{q^2}{72}-\frac{m^2}{4} +
  \frac{1}{2}\br{m^2-\frac{q^2}{6}}L_\Lambda+\frac{1}{3} \br{\frac{q^2}{4}-m^2}L\right\}+\nn \\
 &+q_\mu q_\nu\brf{\frac{1}{3}L_\Lambda-\frac{5}{9}+\frac{1}{3}\br{m^2-q^2}L}, \label{eq.TensorIntegral}
}
where
\eq{
L_\Lambda=\log\frac{\Lambda^2}{m^2},  \qquad L=L(\tau)=\int\limits_0^1d x \log(1+x(1-x)\tau), \qquad \tau=-\frac{q^2}{m^2}.
\nn
}
By contracting indices in the tensor integral (\ref{eq.TensorIntegral}), we obtain:
\eq{
    \int\frac{k^2 d k}{A B}=-\Lambda^2-\frac{q^2}{2}-m^2+2m^2L_\Lambda-m^2L.
}
According the renormalization procedure, we can omit terms having the form
 $a q^2+b m^2$ and $(c q^2+d m^2)L_\lambda$.
%Besides we can restrict our consideration by only
%tensor structure containing $g_{\mu\nu}$.

Let us consider now the general integral of the form
\eq{
I_{\mu\nu}=\int\frac{k_\mu k_\nu d k }{(k^2-m_1^2)((k-q)^2-m_2^2)}.
%=A(\tau_1,\gamma)q^2g_{\mu\nu}+O(q_\mu q_\nu),
%\qquad
%\tau_1=-\frac{q^2}{m_1^2}, \qquad \gamma=\frac{m_2^2}{m_1^2}.
}
Now, let us use the following algebraic identity:
\eq{
\frac{1}{(q-k)^2-m_2^2}=\frac{1}{k^2-m_2^2}+\frac{2qk-q^2}{(k^2-m_2^2)^2}+\frac{(2qk-q^2)^2}{(k^2-m_2^2)^2((q-k)^2-m_2^2)}.
}
Due to our renormalization convention, we can omit the first and the second terms in the right-hand side of this equation so the integral reads as:
\eq{
 I_{\mu\nu}=\int\frac{k_\mu k_\nu (2qk-q^2)^2 d k}{(k^2-m_1^2)(k^2-m_2^2)^2((k-q)^2-m_2^2)}.
}
First, we combine the factors $(k^2-m_1^2)$ and $(k^2-m_2^2)^2$ in the denominator using the Feynman trick:
\eq{
    \frac{1}{a^2b}=2\int\limits_0^1 \frac{(1-x)dx}{\br{a(1-x)+bx}^3}
}
and obtain
\eq{
\frac{1}{(k^2-m_1^2)(k^2-m_2^2)^2} =2\int\limits_0^1\frac{(1-x) d x}{(k^2-M_x^2)^3}, \qquad M_x^2=(1-x)m_2^2+x m_1^2.
\nn
}
Next, we join the resulting expression with the factor $((k-q)^2-m_2^2)$ with the similar Feynman identity:
\eq{
    \frac{1}{c^3d}=3\int\limits_0^1 \frac{(1-y)^3 d y}{\br{c(1-y)+d y}^4}.
}
and, finally, get:
\eq{
\frac{1}{(k^2-M_x^2)^3((q-k)^2-m_2^2)} & =3\int\limits_0^1\frac{(1-y)^2 d y}{\br{(k-y q)^2-m_1^2 d}^4},
}
where
\ga{
    d=\tau_1 y(1-y)+\mu^2,
    \qquad
    \mu^2=x(1-y)+\gamma[y+(1-x)(1-y)],
    \nn\\
    \tau_1=-\frac{q^2}{m_1^2},
    \qquad
    \gamma=\frac{m_2^2}{m_1^2}.
    \nn
}
Thus, we have the logarithmically-divergent loop momentum integral, which allows the operation of
the loop momentum shifting $k=\bar{k}+q y$. After that, we can use the loop integrals from the beginning of this Appendix.
Now, we have:
\eq{
 I_{\mu\nu}&=A(\tau_1, \gamma)\ q^2g_{\mu\nu}+O(q_\mu q_\nu), \nn \\
 A(\tau_1, \gamma)&=-\int\limits_0^1 d x\int\limits_0^1 d y (1-x)(1-y)^2 \br{\log d-\frac{\tau_1(1-2y)^2}{2 d}},
}
therefore the renormalization procedure for this integral has the form:
\eq{
    \tau_1 A(\tau_1,\gamma) \to \tau_1 (A(\tau_1,\gamma)-A(-1,\gamma))+(1+\tau_1)A'(-1,\gamma),
}
where $A(-1,\gamma) \approx -0.0896$ and $A'(-1,\gamma) \approx 0.00654$ for
$\gamma = m_H^2/m_Z^2 = 1.879$.

% ======================================================================
%\bibliographystyle{D:/Physics/Bibliography/Styles/h-physrev5}
%\bibliography{D:/Physics/Bibliography/Main}

% =========================================================================
\end{document}